%% file: paper.tex
\documentclass{INTERSPEECH2023}
\usepackage{amsmath,graphicx}
\usepackage{multicol}
\usepackage{multirow}
\usepackage{booktabs}
\usepackage{subfigure}
\usepackage{float}

\interspeechcameraready 
\title{DSE-TTS: Dual Speaker Embedding for Cross-Lingual Text-to-Speech}
\name{Sen Liu, Yiwei Guo, Chenpeng Du, Xie Chen, Kai Yu$^{\dagger}$
\thanks{Kai Yu$^{\dagger}$ is the corresponding author.}}

\address{MoE Key Lab of Artificial Intelligence, AI Institute\\
X-LANCE Lab, Department of Computer Science and Engineering\\Shanghai Jiao Tong University, Shanghai, China}
\email{\{sen.liu, cantabile\_kwok, duchenpeng, chenxie95, kai.yu\}@sjtu.edu.cn}

\begin{document}
\maketitle
\begin{abstract}
Although high-fidelity speech can be obtained for intralingual speech synthesis, cross-lingual text-to-speech (CTTS) is still far from satisfactory as it is difficult to accurately retain the speaker timbres~(i.e. speaker similarity) and eliminate the accents from their first language~(i.e. nativeness). In this paper, we demonstrated that vector-quantized~(VQ) acoustic feature contains less speaker information than mel-spectrogram. Based on this finding, we propose a novel dual speaker embedding TTS (DSE-TTS) framework for CTTS with authentic speaking style. Here, one embedding is fed to the acoustic model to learn the linguistic speaking style, while the other one is integrated into the vocoder to mimic the target speaker's timbre. Experiments show that by combining both embeddings, DSE-TTS significantly outperforms the state-of-the-art SANE-TTS in cross-lingual synthesis, especially in terms of nativeness.
\end{abstract}
\noindent\textbf{Index Terms}: cross-lingual text-to-speech, dual speaker embedding, vector-quantized acoustic feature

\input{content/0_introduction}
\input{content/1_method}
\input{content/2_experiments_results}
\input{content/3_conclusions}
\input{content/4_acknowledgements}

\bibliographystyle{IEEEtran}
\bibliography{citations/res}
\end{document}

%% file: content/0_introduction.tex
\begin{figure*}[t]
\centerline{\includegraphics[width=20.cm]{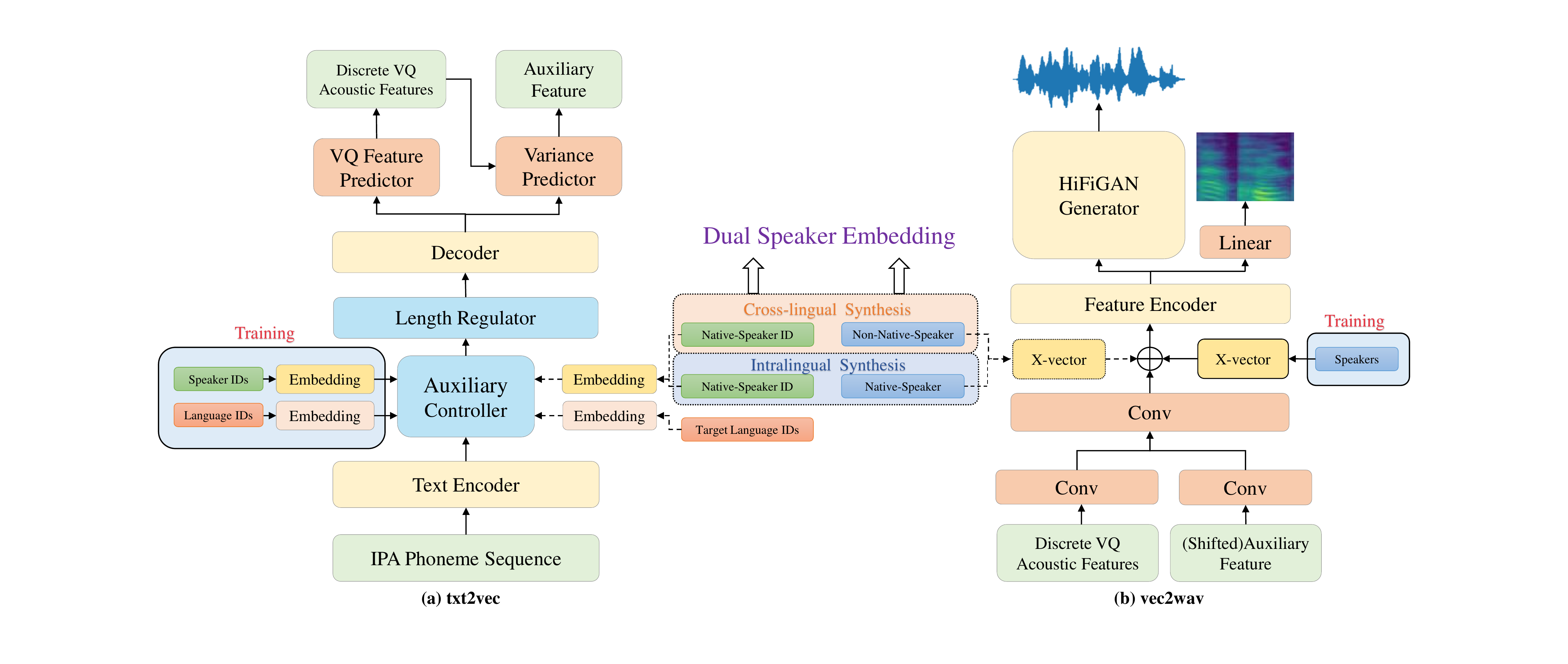}}
\caption{The overall architecture of DSE-TTS consists of an acoustic model, txt2vec, and a vocoder, vec2wav. Dual Speaking Embedding enables DSE-TTS to model linguistic characteristics and speaker timbre separately, helping to retain the speaker's timbre in synthesized speech and eliminating the accents of their native language in cross-lingual synthesis. Solid lines represent the training stage, while dashed lines represent the inference stage.}
\label{fig:framework}
\end{figure*}	
\vspace{-5pt}
\section{Introduction}
\label{sec:intro}
Recent neural text-to-speech (TTS) models \cite{tacotron2,FastSpeech2,RichProsody,emodiff,chenbo} have made great strides in synthesizing speech with high fidelity, rich prosody and remarkable speaker similarity. 
Nevertheless, in multilingual TTS (MTTS) scenarios, cross-lingual synthesis is still far from satisfactory as it is difficult to accurately retain the speaker's timbres and eliminate the accents from their first language. More specifically, cross-language synthesis is difficult to acquire nativeness in non-native languages while maintaining speaker similarity, while nativeness refers to the closeness of speech to the native language.
Efforts have been made to mitigate the degradation in cross-lingual performance resulting from this entanglement.
\cite{learntospeak, DATtts} incorporated adversarial domain training, allowing them to transfer distinct voices across languages. \cite{mutualinfor} proposed to use mutual information minimization to maintain speaker consistency in cross-lingual synthesis. \cite{clTTS} implemented multi-task learning and joint training with a speaker classifier to enhance the overall similarity of speakers. More recently, SANE-TTS \cite{sane-tts} presented an end-to-end multilingual TTS model based on VITS \cite{vits}. This model employed a speaker regularization loss to encourage the model to learn speaker representation independent of its language, ensuring accurate duration predictions in cross-lingual synthesis. 

However, these studies typically rely on the mel-spectrogram as an acoustic feature, which is highly correlated along both time and frequency axes and contains rich speaker-dependent information, making it still challenging to disentangle correlated factors. Seeking another acoustic feature that contains less speaker identity might be crucial.
Recent advancements in speech-based self-supervised learning (SBSSL) \cite{vq-wav2vec,wav2vec2,XLSR53,HuBERT,data2vec,review} have enabled some TTS models to use discrete vector-quantized (VQ) speech representations as an acoustic feature, replacing the traditional mel-spectrogram for prediction. SBSSL models take raw waveform as input, which is only correlated along the time axis. As a result, the quantized output has a coarser granularity of speech features than the mel-spectrogram. This results in lower reconstruction difficulties of VQ features and potentially less speaker-dependent information. For example, \cite{vqtts} leverages self-supervised VQ acoustic features as an alternative to the mel-spectrogram. The VQ features are generated by an acoustic model named txt2vec and then used for waveform reconstruction by a vocoder, vec2wav. By replacing the mel-spectrogram regression task with a VQ feature classification task, \cite{vqtts} achieves highly competitive naturalness among publicly available TTS systems. 

In this paper, by analysizing the performance of the Multi-speaker version of \cite{vqtts}, we found that VQ acoustic feature contains little speaker-specific information. Subsequently, we conducted speaker classification experiment using different acoustic features as shown in section 3. Results show that self-supervised VQ features extracted from wav2vec 2.0 \cite{wav2vec2} contain much less speaker-specific information than mel-spectrogram and other candidates. 
Hence VQ features are easier to decouple timbre and linguistic information than traditional mel-spectrogram. 
Based on this finding, we propose DSE-TTS, a TTS system with dual speaker embedding for cross-lingual TTS that enables the system to model linguistic speaking style and speaker timbre separately. 
The dual speaker embedding operates by controlling different speech aspects in the acoustic model and vocoder separately in the inference stage.
Experiments show that by combining both embeddings, DSE-TTS outperforms the state-of-the-art SANE-TTS in both intralingual and cross-lingual synthesis, especially in terms of nativeness.
We will elaborate on the proposed methods and detailed experimental results in later sections.

%% file: content/1_method.tex
\vspace{-5pt}
\section{Dual Speaker Embedding TTS}
\label{sec:method}
The {\em dual speaker embedding TTS} (DSE-TTS) is introduced in detail in this section, input representations, acoustic model architecture, and the proposed dual speaker embedding. The overall framework is shown in Figure 1.
\vspace{-5pt}
\subsection{Input representations}
Following \cite{learntospeak} and \cite{UnifyandConquer}, the input text is initially normalized and converted to International Phonetic Alphabet (IPA) phonemes using the phonemizer \cite{Phonemizer} toolkit. To facilitate alignment between the text and speech, we preserve the tones and stresses of different languages in our input sequences. We also use shared punctuation tokens across languages, categorized into four groups based on pause length, denoted as `sp1', `sp2', `sp3', and `sp4'. Additionally, we use `sil' as the starting and ending token for each sentence. Prior to feeding the input sequence to the text encoder in txt2vec, each phoneme (token) is assigned a 384-dimensional vector using an embedding table.
\vspace{-7pt}
\subsection{Model architecture}
\vspace{-3pt}
\subsubsection{Self-supervised VQ features}
\label{subsec:ssl_vq}
In this study, we extract VQ acoustic features using a wav2vec 2.0 model with two quantized codebooks, each containing 320 codewords. The wav2vec 2.0 model was pre-trained on 10,000 hours of Mandarin data. It quantizes each input speech into multiple frames with a 20ms stride, and each frame can be represented by concatenating two 256-dimensional codewords from each codebook. All possible index combinations in our mixed-language dataset are about 28.8k. The objective is to accurately predict these index pairs in order to construct high-fidelity speech. For parallel inference, we replace the VQ feature predictor with a convolutional neural network instead of LSTM \cite{LSTM} in the original \cite{vqtts}. Furthermore, we predict the index of each codebook separately instead of their combinations, resulting in two 320-class classification problems. We choose wav2vec 2.0 as our VQ feature extractor because it provides a more robust speech representation with less speaker information compared to other VQ features. The rationale for this choice will be further explained in the experiment section.
\vspace{-5pt}
\subsubsection{Phone-level(PL) auxiliary labelling}
Similar to \cite{vqtts}, we utilize log pitch, energy, and probability of voice (POV)\cite{POV} as auxiliary features. To begin, we compute and normalize phone-level representations of our mixed-language dataset. Then, we apply k-means clustering to group these representations into 128 distinct classes, with the resulting clustered index serving as the auxiliary label for PL information. We employ the ground truth PL auxiliary labels on one side for training the multilingual auxiliary controller, while on the other side, they serve as a condition for subsequent duration modeling and acoustic feature generation.
\vspace{-5pt}
\subsection{Dual speaker embedding}  
In previous works on cross-lingual TTS, it is difficult to accurately retain the speaker's timbres and eliminate the accent from their first languages, resulting in unnatural synthesized speech. The main reason is typically rooted in the entanglement between speakers and languages, which is often manifested in the nature of traditional acoustic features like mel-spectrogram. However, our preliminary experiments found that self-supervised VQ features contain much less speaker identity compared to traditional acoustic features. We will show these results in section \ref{subsec:si-vq}. As a result, in VQ-based TTS methods, it is unnecessary to use additional techniques for the disentanglement of speaker and language within the acoustic model. This allows the model to concentrate solely on modeling textual and linguistic characteristics while the task of controlling speaker timbre is delegated to the vocoder. Thus, the VQ-based TTS model naturally learns how to speak different languages in a native way with the timbre of a non-native speaker. 

From this perspective, we develop a dual speaker embedding TTS (DSE-TTS) framework to improve the nativeness and speaker similarity in cross-lingual TTS scenarios. Two speaker embeddings are used in the TTS model, where one is fed into the acoustic model txt2vec and the other for the vocoder vec2wav. In the training stage, given the text and speech pair of a native speaker, the two speaker embeddings both correspond to the same speaker.
Then in the synthesis stage, no matter the intralingual or cross-lingual case, the speaker embedding of a native speaker corresponding to the input language is chosen as the input speaker embedding to txt2vec uses a native speaker in the language of input text. 
In contrast, the speaker embedding in vec2wav is set as the target speaker.
Hence, in the cross-lingual case, it means that we choose a native speaker's embedding in txt2vec representing linguistic speaking style and the target speaker's embedding in vec2wav that controls the timbre.
In this way, language-specific speaking style and speaker timbre are naturally separated by dual embeddings.

The diagram of DSE-TTS is shown in Figure \ref{fig:framework}. For the acoustic model txt2vec, we take speaker and language IDs as input. Speaker IDs are embedded in 256-dimensional vectors, which are then projected and added to the encoder output. 
We handle language IDs similarly to support various languages, which are embedded in 128-dimensional vectors. These two embeddings are used to learn the linguistic characteristics of different languages. For vec2wav, we use X-vector \cite{xvector} as the speaker embedding to control the timbre, which is extracted from a pre-trained speaker recognition model. 
Besides, to bring the timbre closer to the target speaker while doing cross-lingual synthesis, we shift the distribution of the native speaker's pitches predicted by txt2vec to match the pitch of the target speaker. It can be formulated as follows:
\begin{equation}
    P_{\text{tgt}}=\sigma_{\text{tgt}} \frac{P_{\text{ntv}}-\mu_{\text{ntv}}}{\sigma_{\text{ntv}}} + \mu_{\text{tgt}}
\end{equation}  
where the subscripts ``tgt" and ``ntv" stands for target and native speaker, respectively. $\mu$ and $\sigma$ are the mean and standard deviation of the target or native speaker's pitch values in the training set. We perform this pitch shift before the auxiliary features are sent to vec2wav for synthesis.

%% file: content/2_experiments_results.tex
\vspace{-3pt}
\section{Experiments and results}
\subsection{Dataset}
Our dataset comprises four languages: Mandarin (ZH), English (EN), Spanish (ES), and German (DE). We obtained the data for German and Spanish from M$\_$AILABS \cite{mailabs}, while the data for English and Mandarin were sourced from LibriTTS \cite{LibriTTS} and Aishell3 \cite{aishell3}, respectively. In reality, it may be hard for some languages to collect enough data. To imitate this scenario and test our method's language adaptive ability, we randomly selected a few hours of data from German and Spanish as low-resource languages. The total duration and number of speakers involved are listed in Table \ref{table:dataset}. 
During training, we resampled all speech to 24 kHz and used 5$\%$ of the utterances for the validation and test set. To extract ground truth phoneme duration, we employed MFA\footnote{\url{https://github.com/MontrealCorpusTools/Montreal-Forced-Aligner}}, which performs forced alignment using Kaldi \cite{kaldi}.
\vspace{-2pt}
\begin{table}[H]
 \centering
 \caption{Details of the training dataset.}
 \vspace{-5pt}
 \label{table:dataset}
 \resizebox{0.7\columnwidth}{!}{
 \small
 \begin{tabular}{lcccc}\toprule
    Language   & EN  & ZH & DE    & ES  \\\midrule
    Hours    & 74 & 60 & 6 & 6 \\
    \#Speakers & 228 & 142 & 3 & 3
    \\\bottomrule
 \end{tabular}
 }
\end{table}
\vspace{-5pt}
\subsection{Experimental setup}
We trained our models for 200 epochs on the txt2vec and 100 epochs on the vec2wav, using batch sizes of 16 and 8, respectively. The training process was performed separately on an NVIDIA 2080Ti GPU. We utilized a publicly available pre-trained wav2vec 2.0 model \footnote{\url{https://github.com/TencentGameMate/chinese_speech_pretrain}} for VQ acoustic feature extraction. Additionally, we adopted the data balance strategy proposed by \cite{TowardsUT}, with the scaling factor set to 0.2. To evaluate the performance of our model, we used the recent MTTS model, SANE-TTS, as our baseline and replicated it using the official VITS\footnote{\url{https://github.com/jaywalnut310/vits}} implementation. We trained the SANE-TTS model for 200 epochs using a batch size of 16 while keeping all other parameters consistent with those specified in the original paper.
\vspace{-14pt}
\subsection{Speaker-independent VQ acoustic feature}
\label{subsec:si-vq}
\begin{figure}[hb]
    \centering
    \includegraphics[width=0.8\linewidth]{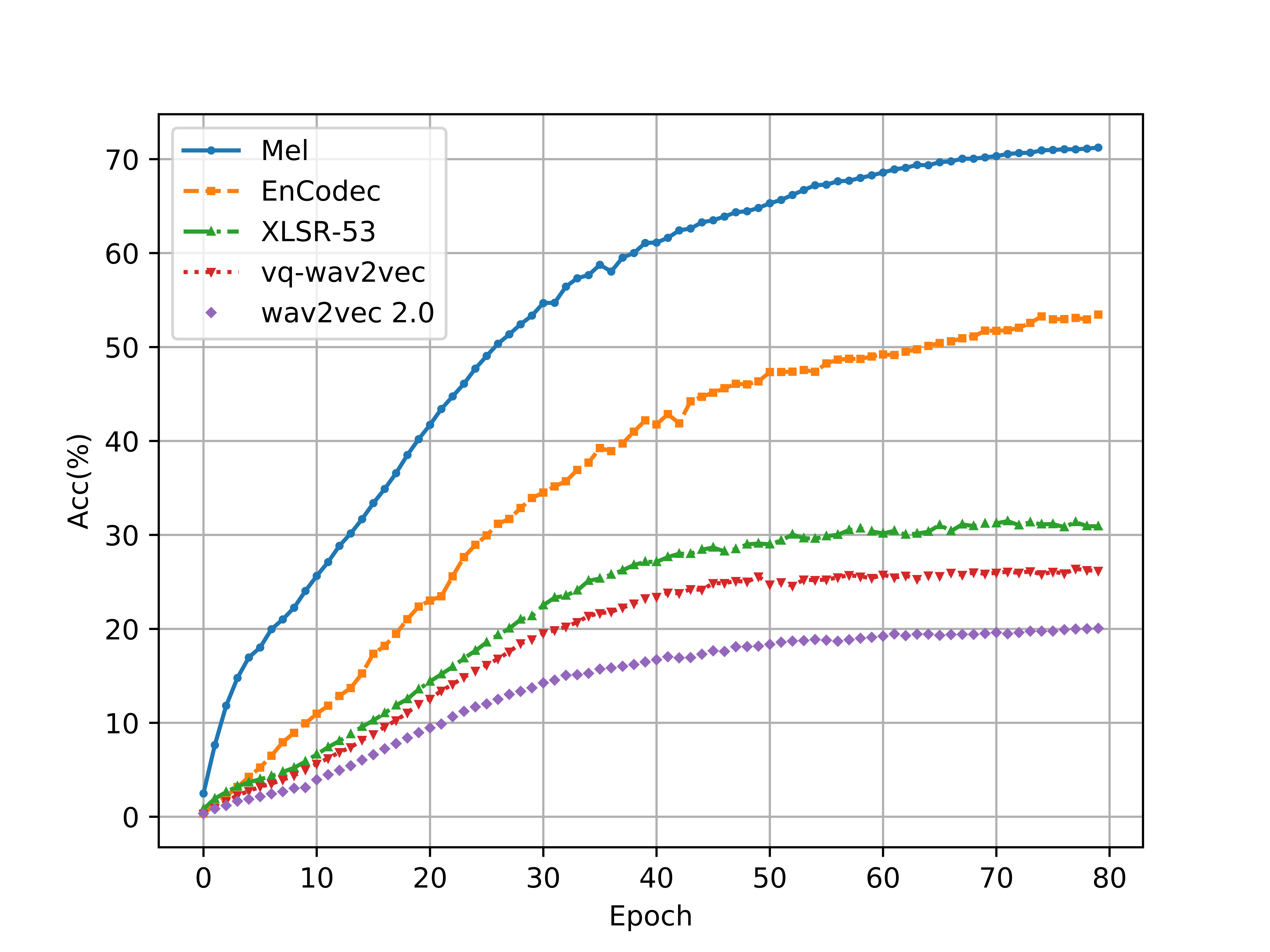}
    \vspace{-0.3cm}
    \linespread{0.9} 
    \caption{Speaker classification accuracy with
different acoustic features.}
    \label{fig:cls_res}
\vspace{-0.5cm}
\end{figure}

To investigate the relationship between different acoustic features and speakers, we first construct a speaker classification model to evaluate the classification accuracy of various features. We compared the mel-spectrogram, a widely used acoustic feature in TTS models, with four distinct VQ features extracted from open-sourced pre-trained models, including vq-wav2vec \cite{vq-wav2vec}, wav2vec 2.0 \cite{wav2vec2}, XLSR-53 \cite{XLSR53} and Encodec \cite{encodec}. 
Our classification model used an X-vector architecture augmented with two linear layers to predict speaker identities. We trained the model on the LibriTTS training set, which includes more than 2000 speakers. After training the model for 80 epochs, we analyze the classification accuracy of speaker identities on the test set. As shown in Figure \ref{fig:cls_res}, the mel-spectrogram contains sufficient information about the speaker's identity, resulting in a high accuracy rate for speaker classification. In contrast, the VQ features have significantly less speaker information, leading to a lower accuracy rate than the mel-spectrogram.
Based on our experimental results, we chose wav2vec 2.0 as our acoustic feature because it has a relatively lower speaker identification performance, indicating that it contains less speaker-dependent information. 
\vspace{-12pt}
\vspace{-5pt}

\begin{table*}[t]
 \centering
 \caption{WER, Speaker Similarity, Nativeness MOS and Similarity MOS results in cross-lingual synthesis, where SECS means Speaker Embedding Cosine Similarity and "w/o DSE" means without Dual Speaker Embedding (a.k.a., using the same target speaker's embedding in both the txt2vec and vec2wav modules during inference).}
 \label{table:cross_eva}
 \resizebox{\textwidth}{!}{
 \begin{tabular}{llcccccccc}\toprule
    \multirow{2}*{Language} & \multirow{2}*{Model} & \multicolumn{4}{c}{EN Speaker} & \multicolumn{4}{c}{ZH Speaker}\\\cmidrule(lr){3-6}\cmidrule(lr){7-10}
    ~ & ~ &  WER $\downarrow$ & SECS $\uparrow$ & NMOS $\uparrow$ & SMOS $\uparrow$ & WER $\downarrow$ & SECS $\uparrow$ & NMOS $\uparrow$ & SMOS $\uparrow$ \\\midrule
    \multirow{3}*{DE} & \text{SANE-TTS} & $29.6$ & $0.44$ & $3.77\pm0.07$ & $4.36\pm0.08$ & $29.6$ & $0.57$ & $3.79\pm0.06$ & $4.54\pm0.08$\\
    & \text{Ours w/o DSE} & $16.1$ & $0.50$ & $4.03\pm0.06$ & $4.45\pm 0.08$ &  $22.6$ & $0.60$ & $3.93\pm0.05$ & $4.60\pm 0.07$\\
    & \textbf{Ours (DSE-TTS)} & $\mathbf{14.9}$ & $0.46$ & $\mathbf{4.19}\pm \mathbf{0.07}$ & $4.40\pm 0.06$ & $\mathbf{15.5}$ & $0.59$ & $\mathbf{4.11}\pm \mathbf{0.06}$ & $4.54\pm 0.07$ \\\midrule
    \multirow{3}*{ES} & \text{SANE-TTS}  & $17.9$ & $0.46$ & $4.03\pm0.06$ & $4.54\pm0.07$ & $21.4$ & $0.57$ & $3.93\pm0.06$ & $4.53\pm0.08$\\
    & \text{Ours w/o DSE} & $17.4$ & $0.49$ & $4.19\pm0.05$ & $4.59\pm 0.08$ & $18.3$ & $0.58$ & $3.97\pm0.06$ & $4.57\pm 0.08$\\
    & \textbf{Ours (DSE-TTS)}& $\mathbf{13.5}$ & $0.50$ & $\mathbf{4.47}\pm \mathbf{0.06}$ & $4.50\pm 0.07$ & $\mathbf{16.0}$ & $0.55$ & $\mathbf{4.26}\pm \mathbf{0.07}$ & $4.52\pm 0.07$ \\\bottomrule
 \end{tabular}
 }
\end{table*}
\vspace{-3pt}

\subsection{Speech Synthesis Evaluation}
We utilized subjective and objective measures to evaluate the quality of intralingual and cross-lingual synthesis. 
Our subjective measures include nativeness mean opinion score (NMOS) and similarity MOS (SMOS). NMOS is used to evaluate the nativeness of synthetic speech, while SMOS is used to assess the extent of speaker similarity. A higher NMOS score indicates that the synthesized speech is closer to the native language. MOS ratings were based on a 1-5 scale with 0.5-point increments and 95$\%$ confidence intervals. 
We synthesized 30 speech samples for each language using random texts from the test set and recruited multiple raters for evaluation. The raters included 15 bilingual Mandarin and English speakers to assess the quality of English and Mandarin speech and 15 trilingual Mandarin-English-German and English-Mandarin-Spanish speakers to evaluate the synthesized speech in German and Spanish, respectively.
For the objective metrics, we computed word error rate (WER), character error rate (CER), and speaker embedding cosine similarity (SECS) between the synthesized speech and the ground-truth speech. WER was used for Spanish, German, and English, while CER was used for Mandarin. We used pre-trained ASR models, Whisper\cite{whisper} for Spanish, German, and English, and a transformer\cite{transformer} ASR model for Mandarin. 
For speaker similarity, we used an independently trained ResNet-based r-vector speaker verification model \cite{r-vector} and computed cosine similarity scores between 0 and 1. A larger score indicates better speaker similarity. To compare our proposed and baseline models, we synthesized 100 speech samples per language by randomly selecting sentences from the test set. Audio samples are available online\footnote{\url{https://goarsenal.github.io/DSE-TTS}}.
\vspace{-8pt}
\begin{table}[H]
 \centering
 \caption{Nativeness MOS and ASR results in intralingual synthesis, while WER is for German (DE), Spanish (ES), and English (EN), and CER is for Mandarin (ZH).}
 \vspace{-6pt}
 \label{table:intra_eva}
 \resizebox{\columnwidth}{!}{
 \begin{tabular}{llcc}\toprule
    Language & Model & NMOS $\uparrow$ & WER(CER) $\downarrow$ \\\midrule
    \multirow{3}*{DE} & \text{Ground truth} & $4.49\pm0.07$ & $6.4$ \\
    & \text{SANE-TTS} & $4.08\pm0.08$ & $16.4$\\
    & \textbf{Ours (DSE-TTS)} & $\mathbf{4.40}\pm \mathbf{0.07}$ & $\mathbf{7.8}$ \\\midrule
    \multirow{3}*{ES} & \text{Ground truth} & $4.69\pm0.06$ & $4.1$ \\
    & \text{SANE-TTS} & $4.30\pm0.07$ & $9.2$\\
    & \textbf{Ours (DSE-TTS)} & $\mathbf{4.56}\pm \mathbf{0.05}$ & $\mathbf{8.8}$ \\\midrule
    \multirow{3}*{EN} & \text{Ground truth} & $4.54\pm0.05$ & $4.2$ \\
    & \text{SANE-TTS} & $4.18\pm0.06$ & $5.6$\\
    & \textbf{Ours (DSE-TTS)} &  $\mathbf{4.36}\pm \mathbf{0.06}$ & $\mathbf{5.3}$ \\\midrule
    \multirow{3}*{ZH} & \text{Ground truth} & $4.46\pm0.06$ & $6.8$ \\
    & \text{SANE-TTS} & $3.79\pm0.07$ & $10.6$\\
    & \textbf{Ours (DSE-TTS)} & $\mathbf{4.39}\pm \mathbf{0.06}$ &  $\mathbf{7.9}$ \\\bottomrule
 \end{tabular}
 }
\end{table}
\vspace{-20pt}
\subsubsection{Intralingual synthesis}
\vspace{-3pt}
Table \ref{table:intra_eva} shows the average NMOS and WER (CER) in intralingual evaluation. It is evident that DSE-TTS has achieved NMOS scores close to the ground truth and outperforms the baseline model on all metrics and across all languages. Specifically, DSE-TTS has attained an NMOS score above 4.3 for each language and achieved a lower WER (CER).
\vspace{-5pt}
\subsubsection{Cross-lingual synthesis}
\vspace{-3pt}
Table 2 presents the evaluation results of our cross-lingual synthesis. We observe that the results were consistent with those obtained in intralingual synthesis, as DSE-TTS outperformed SANE-TTS in terms of both NMOS and WER scores by a large margin. 
Specifically, in NMOS scores, raters preferred DSE-TTS against baseline by over 0.3 in all the speaker-language combinations.
Moreover, the SMOS and SECS scores demonstrate that DSE-TTS maintains similar speaker characteristics to SANE-TTS. These findings suggest that DSE-TTS can synthesize high-quality German and Spanish speech in a non-native speaker's voice but with greater similarity to that of native speakers than the baseline model. 
\vspace{-8pt}
\subsubsection{Ablation study}
\vspace{-3pt}
We performed an ablation study to investigate the impact of dual speaker embedding (DSE) on the performance of our model. The results presented in Table \ref{table:cross_eva} indicate a significant enhancement in the nativeness and decrease in WER of the synthesized speech after the integration of DSE. Our observations also suggest that the use of DSE resulted in a slight decrease in the speaker similarity scores in comparison to not using it. This may be attributed to the fact that different languages have unique linguistic speaking styles, and non-native speakers may sound slightly different when speaking a foreign language fluently. This is also evidence that DSE-TTS produces speech in a native way, though not trained with bilingual speakers.

%% file: content/3_conclusions.tex
\vspace{-3pt}
\section{Conclusions}
In this paper, we propose DSE-TTS, a cross-lingual TTS model, which consists of a dual speaker embedding to model linguistic speaking style and speaker timbre separately. We first showed by a preliminary study that VQ features have fewer speaker-dependent features. Leveraging this finding, we improved our model with a novel dual speaker embedding, resulting in cross-lingual speech synthesis with high nativeness and a similar timbre to the target speaker. Our experiments demonstrated that DSE-TTS outperforms SANE-TTS in both intralingual and cross-lingual synthesis, particularly in terms of nativeness. We also verified the effectiveness of dual speaker embedding by an ablation study. In future work, we will focus on enhancing the quality of the synthesized speech in cross-lingual scenarios and expand our model into other languages.

%% file: content/4_acknowledgements.tex
\vspace{-12pt}
\section{Acknowledgements}
This study was supported by Shanghai Municipal Science and Technology Major Project (No.2021SHZDZX0102) and the Key Research and Development Program of Jiangsu Province, China (Grant No.BE2022059-1).